\documentclass[aps,pra,twocolumn,amsmath,amssymb,nofootinbib,showpacs,superscriptaddress]{revtex4}
\usepackage[english]{babel}
\usepackage{latexsym}
\usepackage{graphics}
\usepackage[demo]{graphicx}
\usepackage{epsfig}
\usepackage{bm}
\usepackage{amsmath}
\usepackage{amssymb}
\usepackage{amsthm}
\usepackage{dcolumn}
\usepackage{bm}
\usepackage{subfig}
\usepackage{float}
\usepackage{tikz} 
\usetikzlibrary{calc}
\usepackage{pgfplots}
\pgfplotsset{compat=newest, width=8 cm}
\usepgfplotslibrary{external}
\usepgfplotslibrary{groupplots}
\usetikzlibrary{arrows.meta}
\usepgfplotslibrary{fillbetween}
\usepgfplotslibrary{polar}
\usepackage{pgfplotstable}
\usepackage{hyperref}
\graphicspath{{Images/}}
\usepackage{comment}
\usepackage{epstopdf}
\usepackage{cancel}
\usepackage{cleveref}
\usepackage{enumerate}
\usepackage{braket}
\usepackage{appendix}
\usepackage{standalone}
\usepackage{capt-of}
\newcommand{\bs}{\boldsymbol}
\newcommand{\ovl}{\overline}
\DeclareMathOperator\erf{erf}
\DeclareUnicodeCharacter{2009}{{ }}

\theoremstyle{remark}

\pgfplotsset{colormap={violet}{rgb255=(25,25,122) rgb255=(238,140,238) color=(white)}}

\makeatletter
\pgfmathdeclarefunction{erf}{1}{%
  \begingroup
    \pgfmathparse{#1 > 0 ? 1 : -1}%
    \edef\sign{\pgfmathresult}%
    \pgfmathparse{abs(#1)}%
    \edef\x{\pgfmathresult}%
    \pgfmathparse{1/(1+0.3275911*\x)}%
    \edef\t{\pgfmathresult}%
    \pgfmathparse{%
      1 - (((((1.061405429*\t -1.453152027)*\t) + 1.421413741)*\t 
      -0.284496736)*\t + 0.254829592)*\t*exp(-(\x*\x))}%
    \edef\y{\pgfmathresult}%
    \pgfmathparse{(\sign)*\y}%
    \pgfmath@smuggleone\pgfmathresult%
  \endgroup
}
\makeatother

\pgfplotsset{
        compat=1.12,
        /pgf/declare function={
            f(\x) = 1+sqrt(2*\x/pi)*exp(-0.5/(\x))/(1+erf(sqrt(0.5/(\x));
        },
        /pgf/declare function={
            density01(\x) = exp(-(\x-1)*(\x-1)/(2*0.1))/(sqrt(0.5*pi*0.1)*(1+erf(sqrt(0.5/0.1))));
        },
        /pgf/declare function={
            density20(\x) = exp(-(\x-1)*(\x-1)/(2*20))/(sqrt(0.5*pi*20)*(1+erf(sqrt(0.5/20))));
        },
        /pgf/declare function={
            density1(\x) = exp(-(\x-1)*(\x-1)/(2*1))/(sqrt(0.5*pi*1)*(1+erf(sqrt(0.5/1))));
        },
    }

\begin{document}

\title{Stochastic model for assesing coherent properties of polariton condensates}
\author{N.A. Asriyan}
\email{norayras@gmail.com}
\affiliation{N.L. Dukhov Research Institute of Automatics (VNIIA), Moscow 127055, Russia}

\author{A.A. Elistratov}
\affiliation{N.L. Dukhov Research Institute of Automatics (VNIIA), Moscow 127055, Russia}

\author{Yu.E. Lozovik}
\affiliation{Institute for Spectroscopy RAS, Troitsk 108840, Moscow, Russia}
\affiliation{MIEM, National Research University Higher School of Economics, Moscow 101000, Russia}
\begin{abstract}
    By considering a microscopical model, we derive an evolution equation for single-mode polariton condensate taking into account the fluctuations of the order parameter. We use it to derive an analytical expression for the second order correlation function and the share of coherent occupation of the condensate in presence of pumping-leakage balance.
\end{abstract}

\maketitle
\section{Introduction}
Due to the complex nature of excitonic polaritons, with low photonic mass being put along with excitonic interactions, they serve as an excellent experimental platform for investgating bose-condensation. However, due to their finite lifetime, these systems demonstrate multiple effects which are not observed in ultracold atomic condensates and are specific to non-equilibrium condensation. Among them are spontaneous vortex lattice formation~\cite{lagoudakis_quantized_2008}, condensation on a ring in momentum space in small samples~\cite{richard_spontaneous_2005}. The collective excitation spectrum is shown to be diffusive at small momenta~\cite{szymanska_nonequilibrium_2006}.

Most of the effects of that type have been successfully explained using extensions of the well-known Gross-Pitaevskii equation (GPE) by addition of driving and dissipative terms~\cite{keeling_spontaneous_2008, wouters_excitations_2007}. One usually takes advantage of well-defined separation of two regions on the dispersion curve, which correspond to polaritonic condensate and quasi-equilibrated excitonic reservoir (usually at a temperature higher than the lattice one) as shown in Fig. \ref{fig:polariton_curve}. Therefore, condensate evolution is described by an equation of the following type:
\begin{align}\label{eq:GPE}
    i\partial_t \psi(r)=\left[-H_0+\frac{i}2\left(R(n_{\rm R})-\Gamma\right)+g|\psi|^2\right]\psi.
\end{align}
Here $H_0=-\frac{\Delta}{2 m_{\rm pol}}$ is the free-particle Hamiltonian, $\Gamma$ stands for the decay rate of polaritons through the imperfect mirrors of the cavity, and $R(n_{\rm R})$ is a function of reservoir density $n_R$, which is governed by a coupled Boltzmann equation. For fast reservoir relaxation, which is usually the case, it may be considered to be an 
adiabatic variable, following condensate density.

Despite the success of these methods, they are not able of describing statistical properties of the condensate, which are due to fluctuations. Among the essential quantities, which can not be evaluated with these mean-field models are the first, second and even higher order coherence functions~\cite{horikiri_higher_2010}. They are especially informative in case of polaritons since photons, which escape the condensate through the cavity mirror are accessible in experiments. Thus, measurements of coherence functions, especially the time-resolved one, provide an opportunity to study the processes in the polaritonic condensate almost directly. Moreover, coherent properties of polaritonic condensates are expected to be useful resource for quantum processing tasks~\cite{luders_quantifying_2021}.

To develop a model which provides access to statistical properties of the condensate, the reservoir states may be described by the Boltzmann kinetic equation. However, to take into account fluctuations in the condensate, one should go beyond this. Among the proposed theoretical approaches are using the master equation for the condensate~\cite{schwendimann_statistics_2008, sarchi_effects_2008}, utilizing the truncated Wigner approximation (TWA)~\cite{wouters_stochastic_2009} or employing the Keldysh technique~\cite{h_haug_quantum_2014, elistratov_polariton_2018} to derive a stochastic version of the driven-dissipative Gross-Pitaevskii equation. 
The models from \cite{wouters_stochastic_2009} and \cite{schwendimann_statistics_2008} have been the main tools for exploring statistics of polariton condensates. They were succesfully applied to explain multiple experimental observations~\cite{love_intrinsic_2008, horikiri_higher_2010, adiyatullin_temporally_2015, klaas_photon-number-resolved_2018}.

Inspired by these models, the current study is, in some sense, in between the two described approaches (the use of classical GPE model with no fluctuations and elaborate simulations of master equation). We aim to capture the effect of noise terms in a simple manner which still allows analytical investigations. Physically, they originate from interaction of the condensate with the exciton pumping reservoir and the photonic decay bath. The subject of our interest is the impact of stochastic drive on the condensate coherence. We will use the polaritonic system as a motivation to consider a general setup for the problems of this type and some general features of condensate coherence will be described.

We will derive the Langevin evolution equation and use it to determine the coherence degree of the condensate depending on the parameters of the polaritonic system. We will demonstrate that in Markovian evolution regime there exist universal (in a sense that it does not depend on subtle microscopic details of the polaritonic system) analytical single-parameter expressions for the second-order coherence function $g_2(0)$ and condensate mode occupation. We demonstrate the possibility of deriving an equation with the saturation model proposed phenomenologically in~\cite{keeling_spontaneous_2008}, namely:
\begin{align}
    i \hbar \partial_t \psi=\left[-\frac{\hbar^2 \nabla^2}{2 m}+V(r)+U|\psi|^2+i\left(\gamma_{\mathrm{eff}}-\Gamma|\psi|^2\right)\right] \psi
\end{align}
from microscopic considerations. Moreover, we show that with properly added noise terms an equation of this type suits for description of condensate statistics.

The content of the paper in 6 main sections. The first Section \ref{sec:motivation} describes microscopically a polaritonic system of interest, a minimalistic model of the system is presented in the next Section \ref{sec:model}. Then comes the derivation of the stochastic evolution equation (Section \ref{sec:derivation}) followed by analysis of its solution (Section \ref{sec:solution}). A small Section \ref{sec:observables} between the two reveals the connection between the stochastic averages with observable occupation and coherence function.

\section{Motivation: polaritonic system}\label{sec:motivation}
As already mentioned, when describing polaritons under incoherent pumping, the condensate is usually considered as an open system. It is subject to decay due to photon leakage outside the microcavity and is interacting with the excitonic cloud in high-energetic states.

\begin{figure}[htp]
\begin{center}
      \includegraphics{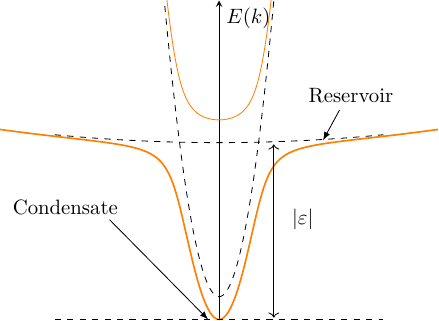}  
\end{center}
    \caption{Schematic of polaritonic spectrum with separation of condensate and reservoir demonstrated.}
    \label{fig:polariton_curve}
\end{figure}

Usually, due to slow cooling rate of excitonic cloud, it is considered to be a quasi-equilibrium reservoir for polaritonic condensate. This is demonstrated schematically in Fig. \ref{fig:polariton_curve}.

We are going to consider a reasonably simplified model of this system by confining the condensate to a single energy level (the zero-momentum state). Moreover, assuming rapid relaxation due to interparticle collisions, the exciton system is assumed to be thermalized at some temperature $T$ and chemical potential $\mu$. These assumptions repeat the ones used in our preceding paper~~\cite{asriyan_mean_2023}.

The Hamiltonian for this system is as follows (hereafter the bar stands for complex conjugation and $\hbar=1$):
\begin{align}\label{eq:Hamiltonian}
    \hat H = &-\varepsilon\hat \psi_0^{\dagger}\hat \psi_0+\sum_{\bs q\neq\bs 0}\varepsilon_{\bs q}\hat \psi_{\bs q}^{\dagger}\hat \psi_{\bs q}+\sum_{\bs n}\omega_n\hat c_{n}^{\dagger}\hat c_{n}+\nonumber\\
    +g_0 &\sum_{\boldsymbol{q}_{1}, \boldsymbol{q}_{2}, \boldsymbol{q}^{\prime}} \hat{\psi}_{\boldsymbol{q}_{1}{+}\bs q^{\prime}}^{\dagger} \hat{\psi}_{\boldsymbol{q}_{2}{-}\boldsymbol{q}^{\prime}}^{\dagger} \hat{\psi}_{\boldsymbol{q}_{1}} \hat{\psi}_{\boldsymbol{q}_{2}} + \sum_{n} \left(t_n\hat\psi_{\bs 0}\hat c_n^{\dagger} + \ovl t_n\hat\psi_{\bs 0}^{\dagger}\hat c_n\right)
\end{align}
Here $\hat\psi_0$ is the condensate polariton annihilation operator, $g_0=\frac{V_0}{2L^2}$ stands for contact interparticle interaction with $V_0$ being the interaction potential and $L^2$ denoting the quantization area. $\varepsilon$ denotes the  energy offset of the condensate level with respect to the reservoir dispersion curve $\varepsilon_{\bs q\to 0}=0$. The third term describes a photon bath to which the condensate is linearly coupled (to model the leakage outside the microcavity).

To study the system, we use the path-integral approach, which provides direct access to polariton condensate field \cite{elistratov_coupled_2016}. By applying the Schwinger-Keldysh technique, we derive an evolution equation of the following type (the derivation is standard, see Appendix A and refs \cite{stoof_field_1999, kamenev_field_2011}):
\begin{align}\label{eq:general_Langevin}
        i\partial_t \phi &= \left({-}\varepsilon{+}\Sigma_{\rm ph}^{\rm local}{+}\Sigma_{\rm exc}^{\rm local}{+}g|\phi|^2\right)\phi{+}\nonumber\\
        {+}&\int_0^t\Sigma_{\rm ph}^+(t{-}t')\phi (t')dt'{+}\int_0^t\Sigma_{\rm 
        exc}^+(t{-}t')\phi (t')dt'{+}\eta{+}\zeta
\end{align}
Here $\Sigma^{\rm ph/\rm exc}$ stand for the self energies with $\Sigma^+$ being the retarded component and $\Sigma^{\rm local}$ -- the time-local ones.

In addition, $\eta$ and $\zeta$ are the noise terms due to polariton interaction with exciton reservoir and due to photon leakage, respectively. Their correlation functions are given by the Keldysh components of the self-energies:
    \begin{align}
        \braket{\ovl\eta\eta'}=\frac{i}{2}\Sigma_{\rm exc}^K(t-t')\\
        \braket{\ovl\xi\xi'}=\frac{i}{2}\Sigma_{\rm ph}^K(t-t').
    \end{align}

\subsection{Polariton decay}
The structure of the terms $\Sigma^{+/K}_{\rm ph}$ depends on multiple details. However, it's common to consider them to be time-local, which implies
\begin{align}
    \Sigma^+_{\rm ph}=-i\frac{\Gamma}{2}\delta(t,t')
\end{align}
with $\Gamma$ being the decay rate of the condensate polaritons.

To deduce the form of the corresponding $\Sigma_{\rm ph}^K$, we impose the following condition: with coupling to the exciton reservoir being absent, the decay should completely devastate the condensate. This means that the following equation
\begin{align}
    i\partial_t \phi = \left(-\varepsilon{-}i\frac\Gamma{2}{+}g|\phi|^2\right)\phi{+}\zeta
\end{align}
should lead to vanishing occupation number (address to Appendix B where the expressions for the observables are derived) $N_0=\braket{|\phi|^2}-\frac12=0$. This straightforwardly implies
\begin{align}\label{eq:photonic_noisecorr}
    \braket{\ovl\zeta\zeta'}=\frac{\Gamma}{2}\delta(t-t').
\end{align}
\section{Model system}\label{sec:model}
\begin{figure}[htp]
\includegraphics{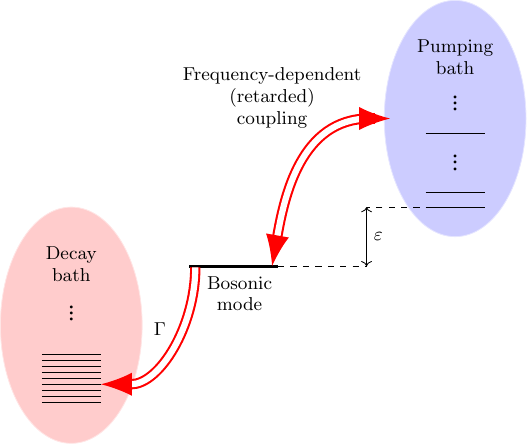}
\caption{Generalization of the polaritonic setup. The isolated bosonic mode is the one where we expect condensate formation due to the coupling with the pumping bath. The latter is considered to have finite memory times, whereas the decay bath is assumed to be Markovian.}
\label{fig:coupling_scheme}
\end{figure}

With the above-presented motivation in mind, we consider a general model of the type shown in Figure \ref{fig:coupling_scheme}.

The evolution of the bosonic field is given by the following equation:
\begin{align}
    i\partial_t \phi &= \left({-}\varepsilon{-}i\frac{\Gamma}{2}{+}g|\phi|^2\right)\phi{+}\int_0^t\Sigma_{\rm 
        pump}^+(t{-}t')\phi (t')dt'{+}\eta{+}\zeta
\end{align}
Here all the local contributions to the energy are assumed to be absorbed into $\varepsilon$.

\subsection{Coupling to the pumping reservoir}
We do not consider any particular form of the components $\Sigma^{+/K}_{\rm pump}$, but only impose two general conditions.

At first, the imaginary part of $\Sigma_{\rm pump}^+(\omega)$, which describes the coherent part of the flux from the reservoir, should be nullified at reservoir chemical potential $\omega=\mu$. This guarantees that the condensate equilibrates with the reservoir (in terms of the chemical potential) in the absence of decay bath. This condition was widely used for deriving model equations for atomic gas condensates in traps~\cite{griffin_bose-condensed_2009, gardiner_kinetics_1997} and may be straightforwardly derived for specific scattering processes for 2D polaritons as we demonstrated in~\cite{asriyan_mean_2023}.

The second condition is the fluctuation-dissipation theorem for the quasi-equilibrated reservoir, which implies:
\begin{align}\label{eq:fdth}
    \Sigma_{\rm pump}^K(\omega){=}2i\Im\left[\Sigma_{\rm pump}^+(\omega)\right]\coth\left(\frac{\omega{-}\mu}{2kT}\right).
\end{align}
Here $\mu$ and $T$ are the effective temperature and chemical potential of the pumping reservoir.

\section{Exploring the evolution}\label{sec:derivation}
\subsection{Condensate state}
Depending on the parameters of the reservoirs/couplings, the bosonic mode may be either in a normal state or in the condensed one. In the former $\braket{\psi}=0$, in the latter $\braket{\psi}\ne0$. Here $\braket{...}$ denotes averaging over the noise distributions. In the condensed state, one may seek the solution of the mean-field evolution equations (neglecting the noise terms) in a form $\phi = \sqrt{\rho_0}e^{-i\nu_0 t}$, which is a valid stationary solution in case the following conditions are satisfied (for details one may address to \cite{asriyan_mean_2023}):
\begin{equation}\label{eq:equilibrium_system}
    \begin{cases}
        \nu_0 = -\varepsilon + g\rho_0+\Re[\Sigma^+(\nu_0)],\\
        \Im[\Sigma^+(\nu_0)]=\frac{\Gamma}2.
    \end{cases}
\end{equation}
Hereinafter we omit the subscript ${}_{\rm pump}$.

The first equation of the system \eqref{eq:equilibrium_system} ensures the establishment of equilibrium chemical potential of the condensate whereas the second describes the balance of particle influx from pumping reservoir and outflux to the decay bath. 

\begin{figure}[htp]
	\centering
	\includegraphics{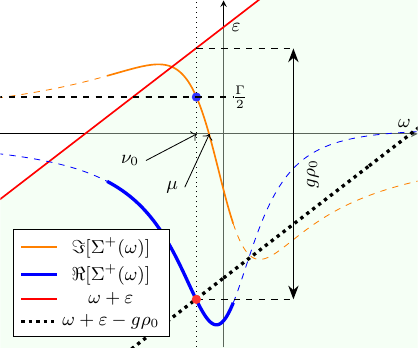}
	\caption{Graphical solution of the equation system \eqref{eq:equilibrium_system}}
	\label{fig:graphsol}
\end{figure}

When considering a stable solution of the system \eqref{eq:equilibrium_system}, we need to impose conditions for stability in a form $\partial_{\rho}\Im[\Sigma^+(\omega)]_{\omega = \nu_0}<0$, which may be expressed as:
\begin{align}\label{eq:stability}
    \frac{g\partial_\omega\left(\Im[\Sigma^+(\omega)]\right)_{\omega = \nu_0}}{1-\partial_\omega\left(\Re[\Sigma^+(\omega)]\right)_{\omega = \nu_0}}<0.
\end{align}
A schematic of the graphical solution of the equation system \eqref{eq:equilibrium_system} is given in Fig. \ref{fig:graphsol}.
Note that we do not need the exact form of the $\Sigma^+(\nu)$ curve. As will be demonstrated, only universal qualitative features in the vicinity of the equilibrium point matter for our study.
\subsection{Stochastic evolution}\label{sec:solution}
Considering the region of the phase space of the system, where the condensed state with $\rho_0\ne0$ is stable, we add the noise terms and use a substitution $\phi(t)=\varphi(t)e^{-i\nu(t)t}$ assuming $\frac{d\ln (\nu(t))}{dt}\ll 1$ (the frequency $\nu$ evolves slowly). This leads to
    \begin{align}
        id_t\varphi &= \left(\varepsilon - \nu(t)-i\frac{\Gamma}2+g|\varphi|^2\right)\varphi+\\
        &+\int_0^t\Sigma^+(t-t')e^{i\nu(t)\times(t-t')}\varphi (t')dt'+\tilde \eta+\tilde \zeta.
    \end{align}
Here $\tilde \eta/\tilde \zeta = \eta/\zeta e^{i\nu(t)t}$.
    
 We are now going to treat $\varphi(t)$ as a slow-evolving function, which allows to transform:
    \begin{align}
        id_t\varphi = \left(\varepsilon {-} \nu(t){-}i\frac{\Gamma}2{+}g|\varphi|^2\right)\varphi{+}\Sigma^+\left[\nu (t)\right]\varphi{+}\tilde \eta{+}\tilde \zeta.
    \end{align}
Here $\Sigma^+[\nu]=\int_{-\infty}^{\infty}\Sigma^+(t)e^{i\nu t}dt$ is the Fourier transform of the retarded self-energy component. One may separate its real and imaginary parts:
    \begin{align}
        id_t\varphi = \left(\varepsilon - \nu(t)\right.&\left.+g|\psi|^2+\Re[\Sigma^+(\nu (t))]\right)\varphi+\nonumber\\
        &+i\left(\Im[\Sigma^+\left(\nu (t)\right)]-\frac{\Gamma}{2}\right)\varphi+\tilde \eta+\tilde \zeta.
    \end{align}

    Now we pass to a rotating frame (where rotation of $\varphi(t)$ is solely due to fluctuations) by seeking $\nu(t)$ as a solution of the following equation:
    \begin{align}
        \varepsilon - \nu(t)+g|\varphi|^2+\Re[\Sigma^+(\nu (t))]=0.
    \end{align}
Since $(\nu_0;\rho_0)$ is a stable solution, we may expand $\nu(t)=\nu_0+\delta \nu$. Using the equilibrium equations \eqref{eq:equilibrium_system}, one may derive (hereafter we utilize shorthand notation $\Re[\Sigma'(\nu_0)]=\partial_\omega\left(\Re[\Sigma^+(\omega)]\right)_{\omega = \nu_0}$, the same for the imaginary part):
    \begin{align}
        \delta \nu(t) = -g(\rho_0-|\psi|^2(t))+\Re[\Sigma'(\nu_0)]\delta\nu(t)\\
        \delta\nu(t) = \frac{g}{1-\Re[\Sigma'(\nu_0)]}\left(|\varphi|^2-\rho_0\right).
    \end{align}
    Expanding the imaginary part, one may express the Langevin equation as follows:
    \begin{align}
        id_t\varphi = i\frac{g\Im[\Sigma'\left(\nu_0\right)]}{1-\Re[\Sigma'(\nu_0)]}\left(|\varphi|^2-\rho_0\right)\varphi+\tilde \eta+\tilde \zeta.
    \end{align}
    Let's denote $\kappa = -\frac{g\Im[\Sigma'\left(\nu_0\right)]}{1-\Re[\Sigma'(\nu_0)]}$.
    Assuming small exciton noise correlation time, we may express the correlation function in a time-local form:
    \begin{align}
        \braket{\ovl{\tilde\eta}\tilde\eta'}=\frac{i}{2}\Sigma^K[\nu(t)]\delta(t-t').
    \end{align}
    Here $\Sigma^K[\nu]$ is the Fourier transform of the Keldysh self-energy component.
    
    Being combined with \eqref{eq:fdth}, this leads to
     \begin{align}
        &\braket{\ovl{\tilde\eta}\tilde\eta'}{=}{-}\Im\left[\Sigma^+(\nu(t))\right]\coth\left(\frac{\nu(t){-}\mu}{2kT}\right)\delta(t{-}t'){=}\\\nonumber
        &{=}{-}\Big(\Im\left[\Sigma^+(\nu_0)\right]\!{+}\Im[\Sigma'(\nu_0)]\delta\nu\Big)\coth\left(\frac{\nu_0{-}\mu{+}\delta\nu}{2kT}\right)\!\delta(t{-}t').
    \end{align}    
    Note that equilibrium equation \eqref{eq:equilibrium_system} allows to substitute:
    \begin{align}
        \braket{\ovl{\tilde\eta}\tilde\eta'}{=}{-}\left[\frac{\Gamma}{2}{+}\Im[\Sigma'\left(\nu_0\right)]\delta\nu\right]\coth\left(\frac{\nu_0{-}\mu{+}\delta \nu}{2kT}\right)\delta(t{-}t').
    \end{align}
    In absence of photonic decay, one may use $\nu_0=\mu$ to expand the correlation function as follows:
    \begin{align}\label{eq:low_gamma}
        \braket{\ovl{\tilde\eta}\tilde\eta'}{=}{-}2\Im[\Sigma'(\nu_0)]kT.
    \end{align}
    This implies that the stability constraint \eqref{eq:stability} is not the only one. We should add the following constraint:
    \begin{equation}\label{eq:noise_positivity}
        \Im[\Sigma'(\nu_0)]<0.
    \end{equation}
    On the contrary, if $\Gamma$ is significant to neglect $\delta\nu(t)$, one deduces the following expression:
    \begin{align}
        \braket{\ovl{\tilde\eta}\tilde\eta'}{=}{-}\frac{\Gamma}2\coth\left(\frac{\nu_0{-}\mu}{2kT}\right)\delta(t-t').
    \end{align}
    Note that due to \eqref{eq:noise_positivity} we may argue that $\nu_0<\mu$ for finite $\Gamma$, which guarantees that $\braket{|\eta|^2}>0$. 
    
    Finally, dealing with two white noise terms $\tilde \eta$ and $\tilde \zeta$, we may treat their sum as a resultant noise term $\xi$ with correlation function (we took advantage here of $\coth(x)$ being an odd function and used \eqref{eq:photonic_noisecorr})
    \begin{align}\label{eq:general_gamma}
        \braket{\ovl\xi\xi'}=\frac{\Gamma}{2}\left[1+\coth\left(\frac{\mu{-}\nu_0}{2kT}\right)\right]
    \end{align}

Actually, this expression reproduces \eqref{eq:low_gamma} when $\Gamma\to 0$, thus we may use \eqref{eq:general_gamma} for all the values of $\Gamma$.

Finally, we expressed the dynamical equation in the following general form:
\begin{align}\label{eq:local_Langevin}
        d_t\psi = \frac{g\Im[\Sigma'\left(\nu_0\right)]}{1-\Re[\Sigma'(\nu_0)]}\left(\rho_0-|\psi|^2\right)\psi+\xi.
\end{align}
We have cancelled the factor of $i$ since $i\xi$ and $\xi$ have the same correlation function.
Equation \eqref{eq:local_Langevin} may be expressed in a form of a complex Ito equation as follows:
\begin{align}\label{eq:complex_Ito}
    d\psi  = \kappa (\rho_0-|\psi|^2)\psi dt + \sqrt{Q}\delta \bs W(t).
\end{align}
Here
\begin{align}
    \kappa &= \frac{g\Im[\Sigma'\left(\nu_0\right)]}{1-\Re[\Sigma'(\nu_0)]},\\
    Q&=\frac{\Gamma}{4}\left[1+\coth\left(\frac{\mu{-}\nu_0}{2kT}\right)\right].
\end{align}
and $\delta \bs W(t)=\delta W_x+i\delta W_y$ is a complex Wiener increment with independent components, \textit{i.e.} $\delta W_x^2{=}\delta W_y^2{=}dt$, $\delta W_x\delta W_y=0$.

Notably, this equation has the from of the one utilized in ~\cite{keeling_spontaneous_2008} (but for a single condensate mode) with properly added noise term.
\section{Observables}\label{sec:observables}
The main goal of the current study is evaluating observables, which may be used to describe the condensed state. These are the occupation number $N$ and the second order coherence function $g_2(0)$. They are given by the following expressions:
\begin{align}
    N &= \braket{\hat \Psi^{\dagger}\hat\Psi}\\
    g_2(0)&=\frac{\braket{\hat \Psi^{\dag}\hat \Psi^{\dag}\hat \Psi\hat \Psi}}{\braket{\hat{N}}^2}.
\end{align}
Note that these expressions are in terms of normal-ordered observables, which, by optical equivalence theorem~\cite{sudarshan_equivalence_1963}, correspond to calculating averages with the Glauber-Sudarshan P-function as a probability distribution. At the same time, the Langevin equation, which we derive by means of the Keldysh technique, naturally samples the Wigner distribution~\cite{stoof_field_1999}. Thus, we need a relation between the average over the noise realizations and the normal ordered ones. They are established in Appendix B with the final results being as follows (here the brackets denote noise distribution averaging):
\begin{align}\label{eq:Langevin_averaged_observables}
    N(t)&=\braket{|\varphi|^2}{-}\frac12,\\
    g_2(0)&=2+\frac{\braket{|\varphi|^4}-2\braket{|\varphi|^2}^2}{\left(\braket{|\varphi|^2}-\frac12\right)^2}.
\end{align}
Here we took into account that $|\phi|^2=|\varphi|^2$.

\section{Solving the evolution equation}\label{sec:solution}
We are going now to utilize \eqref{eq:complex_Ito} and pass to polar coordinates with 
the help of substitution $\phi = \sqrt{\rho}e^{-i\theta}$ and Ito's lemma. This leads to
\begin{equation}\label{eq:radial}
    d\rho = 2\left[\kappa(\rho_0-\rho)\rho+Q\right]dt+2\sqrt{Q\rho}\delta W_r(t),
\end{equation}
\begin{equation}\label{eq:angular}
    d\theta = \sqrt{\frac{Q}{\rho}}\delta W_{\theta}.
\end{equation}
Here $\delta W_r$ and $\delta W_{\theta}$ are independent Wiener increments: $\delta W^2_r=\delta W^2_\theta=dt$.

The angular diffusion, described by \eqref{eq:angular} does not affect the observables of our interest, thus we further focus on the radial one. It's helpful to make it dimensionless by introducing $\tau = \rho_0\kappa t$, $\alpha = \frac{Q}{\kappa \rho_0^2}$ and to pass to dimensionless $\xi=\rho/\rho_0$:
\begin{align}
    d\xi = 2(\xi-\xi^2+\alpha)dt+2\sqrt{\alpha\xi}\omega\sqrt{d\tau}.
    \label{eq:Ito_occup}
\end{align}
The Fokker-Planck equation for the distribution function $P[\xi, \tau]$, corresponding to \eqref{eq:Ito_occup} is as follows:
\begin{align}
    \partial_\tau P=2\alpha \xi \partial^2_{\xi}P+2(\xi^2-\xi+\alpha)\partial_\xi P+2(2\xi-1)P.
\end{align}
Notably, the stationary equation with $\partial_\tau P=0$ is exactly solvable. Of its linearly independent solutions we need the one that nullifies the probability flow:
\begin{align}
    J[\xi, t]=2(\xi-\xi^2)P-2\alpha\xi \partial_\xi P\equiv 0.
\end{align}
Thus, $\partial_\xi[\ln P]=(1-\xi)/\alpha$ and
\begin{align}\label{eq:PDF}
    P[\xi, \infty]=\frac{e^{\frac{-(\xi-1)^2}{2\alpha}}}{\sqrt{\frac{\pi\alpha}{2}}\left(1+\erf{\frac{1}{\sqrt {2\alpha}}}\right)}.
\end{align}
With the help of this distribution function, one may derive an expression for the occupation:
\begin{align}\label{eq:occupation}
    \braket{N}=\rho_0\braket{\xi}-\frac12=\rho_0\left[1+\sqrt{\frac{2\alpha}{\pi}}\frac{e^{-\frac1{2\alpha}}}{1+\erf{\frac{1}{\sqrt {2\alpha}}}}\right]-\frac12,
\end{align}
where $
\braket{
\xi}=\int_0^\infty \xi P[\xi]d\xi{}$.

For the second order correlation function, the expression is as follows:
\begin{align}\label{eq:g2}
    g_2(0)=2+\frac{\alpha+\braket{\xi}-2\braket{\xi}^2}{\left(\braket{\xi}-\frac12\right)^2}.
\end{align}

Not very close to the condensation threshold (where we can neglect the difference between $\braket{N}$ and $\braket{\rho}$), we may introduce $\eta  = 1/{\braket{\xi}}=\rho_0/\braket{\rho}$ as the share of coherent occupation. This is an experimentally accessible observable, \textit{e.g.} see \cite{alnatah_coherence_2024}.

\begin{figure}[htp]
    \includegraphics{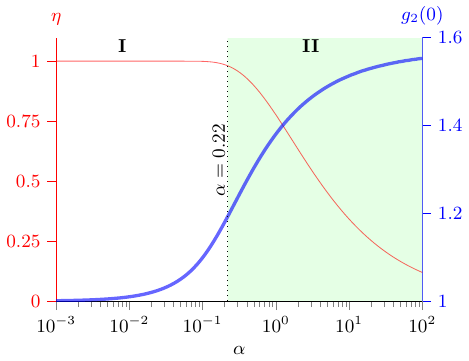}
    \caption{The share of coherent occupation $\eta = \rho_0/\braket{\rho}$ (thin red line, left axis) and the second order coherence function $g_2(0)$ (thick blue line, right axis) as functions of "relative fluctuation strength"\ $\alpha$. The boundaries of regions {I} (zone of partial coherence) and {II} (phase transition region) are given by dotted vertical lines. Note that the region {II} is not bounded from the right.}
    \label{fig:rhog2alpha}
\end{figure}

Under the same assumption, we may simplify:
\begin{align}\label{eq:implicit}
    g_2(0)=\eta+\alpha \eta^2.
\end{align}
Both the variables ($\eta$ and $g_2(0)$) are plotted as functions of $\alpha$ in Fig. \ref{fig:rhog2alpha}.

In the figure one may clearly observe two trends: the increase of $\alpha$ is accompanied by an increase of coherence function $g_2(0)$ and a decrease of the coherent occupation share. Both of the tendencies are intuitive. The parameter $\alpha$ may be treated as the "relative strength of fluctuations"\ and, the higher it is, the more fluctuations contribute to the total occupation and spoil condensate coherence.

Considering the details, we note that for vanishing $\alpha$ condensate is completely coherent ($g_2(0)=1$) with no stochastic contribution to the occupation $(\eta  = 1)$.

For higher $\alpha\lessapprox 0.22$ (region {I} in Figure \ref{fig:rhog2alpha}), the incoherent occupation (due to fluctuations) is still barely present, whereas $g_2$ demonstrates quite a sharp increase up to $\approx 1.22$. This is where the condensate density is still well described by the mean-field model, but the coherence degree is significantly different from unity: the condensate is partially coherent. In contrast, for $\alpha\gtrapprox0.22$ (region {II}), the coherence function $g_2(0)$ slowly increases towards its asymptotic value, while $\eta$ is much stronger dependent on $\alpha$ (asymptotically $\sim 1/\sqrt{\alpha}$). This is actually the phase transition region.

The expressions \eqref{eq:occupation} and \eqref{eq:g2} share a single control parameter $\alpha$, which is non-trivially dependent on the features of the particular system under consideration:
    \begin{align}
    \alpha = \frac12\frac{1-\Re[(\Sigma^+)'(\nu_0)]}{\rho_0}\frac{\Gamma}{2g\rho_0}\frac{1+\coth\left(\frac{\mu{-}\nu_0}{2kT}\right)}{\Im[(\Sigma^+)'\left(\nu_0\right)]}.
\end{align}
It is tempting to explicitly express $g_2$ as a function of $\eta$ by excluding $\alpha$ with the help of \eqref{eq:occupation}. For the limiting cases the task is quite straightforward. Namely, for $\eta\to 0$:
\begin{align}
    g_2(0)&=\frac{\pi}{2}\!-\!(\pi{-}3)\eta\!+\!\frac{\pi^2{-}2\pi{-}4}{2\pi}\eta^2\!+\!\frac{\pi^3{-}4\pi^2{+}8}{\pi^3}\eta^3\!+\!O[\eta^4].
\end{align}
In the opposite limit, for $\eta\to 1$:
\begin{align}
    g_2(0)=\eta+\frac{\eta^2}{W_0\left[\frac1{2\pi(\eta-1)^2}\right]}
\end{align}
Here $W_0[x]$ is the principal branch of the Lambert W-function\footnote{The Lambert W-function is defined as an inverse function of $f(W)=W\exp(W)$.}.

For intermediate values we match the two asymptotics using the following function
\begin{align}\label{eq:explicit_approx}
    g_2(\eta)\!=\!\eta\!+\!\frac{\eta^2}{W_0\left[\frac1{2\pi(\eta\!-\!1)^2}\right]}\!+\!a(\eta\!-\!1)^3\!+\!b(\eta\!-\!1)^4\!+\!c(\eta\!-\!1)^5
\end{align}
with
\begin{align}
    a &= -7+\frac{2}{\pi}-\frac{3\pi}{2}+\frac{1}{W_0\left[\frac1{2\pi}\right]}\approx -3.859,\\
    b &= -12+\frac{4}{\pi}-\frac{3\pi}2+\frac{2}{W_0\left[\frac1{2\pi}\right]}\approx  -1.005,\\
    c &= -5+\frac{2}{\pi}-\frac{\pi}{2}+\frac{1}{W_0\left[\frac1{2\pi}\right]}\approx 1.283.
\end{align}

\begin{figure}
    \centering
    \includegraphics{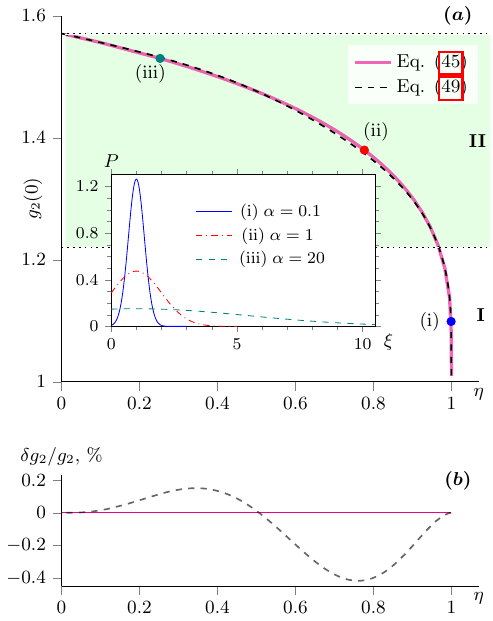}
    \caption{(a) Parametric plot of $g_2(0)$ as a function of condensate fraction $\eta$ (solid line) with an explicit analytical approximation (dashed line). The inset plot shows the distribution function for three values of $\alpha$; (b) Relative approximation error of the implicit dependence, given by \eqref{eq:implicit} with equation \eqref{eq:explicit_approx}.}
    \label{fig:g2rho}
\end{figure}

In Figure \ref{fig:g2rho} (a) we show both the parametric plot $g_2(\eta)$, using \eqref{eq:occupation}-\eqref{eq:g2} and the analytical approximation \eqref{eq:explicit_approx}. The relative approximation error is shown to not exceed  $\leq 0.5\%$ (see panel (b)). Derivation of an accurate expression for $g_2(0)$  as a function of $\eta$ is an important result of our study since it is a prediction with no traces of the specific model under consideration. Hence, it may be directly tested experimentally.

For some values of $\alpha$ the inset plot shows the distribution function \eqref{eq:PDF}. The shapes of the $P[\xi]$ curves explain the origin of the two regions (I and II in Figures \ref{fig:rhog2alpha} and \ref{fig:g2rho}). The distribution function is a truncated Gaussian for the normalized occupation $\xi$. While the fluctuations are weak, the width of the distribution peak increases, whereas its truncated "left tail"\ is negligible. Thus, the mean value $\braket{\xi}$ is barely affected by noise.

For larger values of $\alpha$ in the phase transition region II, the cropped part of the distribution makes the distribution significantly asymmetrical with respect to $\xi=1$, thus $\braket{\xi}$ increases.

Note that even for dominating fluctuations $\alpha\to\infty$, which corresponds to condensation threshold, the asymptotic expression $g_2(0,\alpha = \infty)=\frac{\pi}{2}$ is lower than $2$, which should be the one for the normal state. Clearly, this is due to the fact that we linearized $\nu$ around its "classical"\ value $\nu_0$, which is not a well-defined quantity in the normal phase.

The value $g_2=\pi/2$ at the phase transition threshold is not specific to the polaritonic condensate, the same result applies to the single-mode laser model~\cite{gardiner_93_2010}. Though, the laser equation has a different nonlinear term compared to the one we derived in equation \eqref{eq:complex_Ito}, both equations have the same behavior at threshold. This is due to the fact that one may treat the deterministic part of both the models as a motion along a gradient
\begin{equation}
    \begin{cases}
        \dot x &= -\partial_x V(x, y),\\
        \dot y & = -\partial_y V(x, y)
    \end{cases}
\end{equation}
of a potential of the following form ($\varphi = x+iy$):
\begin{align}
    V(r) = \frac{\kappa\rho_0}{2}|\varphi|^2-\frac{\kappa}{4}|\varphi|^4,
\end{align}
which may be straightforwardly shown to lead to $g_2(0)=\pi/2$ at threshold with $\rho_0=0$~\cite{gardiner_93_2010}. In addition, a similar behavior at condensation threshold may be inferred from numerical simulations of the master equation for the condensate in~\cite{sarchi_effects_2008}.

\section{Conclusion}
Our results support the common notion of condensate coherence behaviour: being subject to radiative decay, the condensate may reach equilibrium in a state with $g_2>1$. We reproduce this behaviour with the help of a simple single-parametric analytic expression and add some details by distinguishing two regions of qualitatively different behaviour of the second order coherence function and the occupation. In addition, we derive an explicit expression of the coherence function as a function of the share of coherent occupation. Since both the quantities we focus on are directly accessible in experiments with polaritons, our prediction may be tested experimentally.

To conclude, we would like to emphasize the general character of the model we studied. Since we did not rely on a particular form of interaction between the pumping reservoir and the condensate, we expect our results to be applicable to wide range of materials and experimental setups where the reason of condensate decoherence is the coupling to a decay bath.

\section{Acknowledgements}
N.A.Asriyan and Yu. E. Lozovik acknowledge the support by the Russian Science Foundation grant No. 23-12-00115, https://rscf.ru/en/project/23-12-00115/.

\newpage
\bibliographystyle{unsrt}
\bibliography{references}
\clearpage
\appendix
\appendixpage
\addappheadtotoc

\label{appendix:A}
\section{Deriving evolution equations for the polariton condensate}
To derive the evolution equation, we utilize the Keldysh-Schwinger technique in path-integral formulation~~\cite{kamenev_field_2011}, which implies the following action for the system, described in the main text:
\begin{align}
    S&{=}\int\limits_c d\tau\left[\ovl{\psi}_0\left\{i\partial_{\tau}{+}\varepsilon\right\}\psi_0{-}\sum_{\bs q\neq \bs 0}\ovl{\psi}_{\bs q}\left\{i\partial_{\tau}{-}\varepsilon_{\bs q}\right\}\psi_{\bs q}\right.\nonumber\\
    &\left.-\sum_n \ovl c_n\left\{i\partial_{\tau}-\omega_n\right\} c_n-\sum_n\left(t_n\psi_{\bs 0}\ovl c_n + \ovl t_n\ovl\psi_{\bs 0} c_n\right)\right.{-}\nonumber\\
    &\left.{-}g_0\sum_{\bs q_1, \bs q_2, \bs q'}\ovl \psi_{\bs q_1+\bs q'}\ovl\psi_{\bs q_2-\bs q'}\psi_{\bs q_1}\psi_{\bs q_2}\right].
\end{align}
By integrating out the reservoir degrees of freedom, one may derive an effective action for the condensate level only. Regardless of the details of the approximations used, the general structure of such an action is as follows (shorthand notation is used $\psi'\equiv \psi(\tau')$):
    \begin{align}
        S^{\rm eff}&{=}\int\limits_c d\tau\left[\ovl{\psi}\left\{i\partial_{\tau}{+}\varepsilon{-}g_0|\psi|^2\right\}\psi{-}\int\limits_c d\tau'\ovl\psi\Sigma(\tau,\tau')\psi'\right].
        \label{eq:condensate_effective_action}
\end{align}
Here we introduced a self-energy term $\Sigma(\tau, \tau')=\Sigma_{\rm exc}(
\tau, \tau')+\Sigma_{\rm ph}(\tau, \tau')$ to describe interaction with both the exciton reservoir and the decay bath.

To describe the real-time dynamics we pass from the field $\psi$ defined on the Keldysh contour (denoted by $c$ in \eqref{eq:condensate_effective_action}) to $\psi_{\pm}$ being the fields on its backward and forward branches. This is achieved by the standard Keldysh rotation
\begin{align}
    \psi_{\pm}=\phi\pm\frac{\xi}{2}.
\end{align}
This procedure makes the real-time action to acquire the following form:
\begin{widetext}
\begin{align}\label{eq:Keldrotated_action}
    &S_{\rm eff}[\phi, \ovl{\phi}, \xi, \ovl{\xi}]=\int_{\tau_0}^{t} d\tau\int_{\tau_0}^{t} d\tau\ovl{\phi}\left[\left\{i\partial_{\tau}{+}\varepsilon{-}\Sigma_{\rm ph}^{\rm local}{-}\Sigma_{\rm exc}^{\rm local}{-}g_0|\phi|^2\right\}\delta(\tau,\tau')-(\Sigma^-_{\rm exc}(\tau,\tau')+\Sigma^-_{\rm ph}(\tau,\tau'))\theta(\tau'{-}\tau)\right]{\xi'}\nonumber\\
    +&\int_{\tau_0}^{t} d\tau\int_{\tau_0}^{t} d\tau'\ovl{\xi}\left[\left\{i \partial_{\tau}{+}\varepsilon{-}\Sigma^{\rm local}_{\rm ph}{-}\Sigma_{\rm exc}^{\rm local}{-}g_0|\phi|^2\right\}\delta(\tau,\tau')-(\Sigma^+_{\rm exc}(\tau,\tau')+\Sigma^+_{\rm ph}(\tau,\tau'))\theta(\tau'{-}\tau)\right]{\phi'}{-}\nonumber\\
    {-}&\frac12\int_{\tau_0}^{t} d\tau\int_{\tau_0}^{t} d\tau'\ovl\xi\left[\Sigma_{\rm exc}^{K}(\tau,\tau')+\Sigma_{\rm ph}^{K}(\tau,\tau')\right]\xi'
\end{align}
\end{widetext}
We have explicitly extracted the time-local (on the Schwinger-Keldysh contour) components of both the exciton and photon contributions to the self-energy term. When deriving the evolution equations from here, the next standard steps are performing the Hubbard-Stratanovich transformation and integrating over $\xi$ and $\ovl\xi$~~\cite{stoof_field_1999}. The result is a Langevin equation of the following form with memory terms, as presented in the main text:
\begin{align}\label{eq:general_Langevin}
        i\partial_t \phi &= \left({-}\varepsilon{+}\Sigma_{\rm ph}^{\rm local}{+}\Sigma_{\rm exc}^{\rm local}{+}g|\phi|^2\right)\phi{+}\nonumber\\
        {+}&\int_0^t\Sigma_{\rm ph}^+(t{-}t')\phi (t')dt'{+}\int_0^t\Sigma_{\rm 
        exc}^+(t{-}t')\phi (t')dt'{+}\eta{+}\zeta
\end{align}
\label{appedix:B}
\section{Deriving an expression for the second-order correlation function}
In this appendix we derive an expression for $g_2(0)$ and occupation number $N$ for the $q=0$ state. At first, we need a relation between the averages obtained by means of Langevin equation and the time-ordered ones of operators. To obtain one, we may consider the following average as a warm-up excercise:
\begin{align}\label{eq:Keldysh_vs_operator}
    \braket{\overline{\psi}(t_1){\psi}(t_2)}=\braket{T_{C}\{\hat \Psi^{\dag}(t_1)\hat \Psi(t_2)\}}.
\end{align}
Here the left-hand side average is over realizations of quantum noise and the right-hand side one is an ensemble Schwinger-Keldysh contour ordered average with initial density matrix $\hat \rho(t_0)$. To relate these averages to the ones obtained from Langevin equation we invert the Keldysh rotation formulas to write:
\begin{align}
&\braket{\phi(t_{1})\overline{\phi(t_{2})}}=\frac1{4}\langle\left[\psi(t^+_{1})+\psi(t^-_{1})\right]\left[\overline{\psi}(t^+_{2})+\overline{\psi}(t^-_{2})\right]\rangle.
\end{align}

Using relation \eqref{eq:Keldysh_vs_operator} for all the four terms, explicitly ordering the terms in time on Schwinger-Keldysh contour and then considering the limit $t_1=t_2$ we obtain:

\begin{align}
    4\braket{|\psi_0|^2}=2(\braket{\hat \Psi\hat \Psi^{\dag}}+\braket{\hat \Psi^{\dag}\hat \Psi}).
\end{align}

After using equal-time commutation relations we get:

\begin{align}
    \braket{|\psi_0|^2}=\braket{\hat{N_0}}+\frac12.
\end{align}

Actually this result may be obtained in another way using $\braket{T_{C}\{\hat \Psi^{\dag}(t_1)\hat \Psi(t_2)\}}=\braket{\{\hat \Psi^{\dag}(t_1)\hat \Psi(t_2)\}}\Theta(t_1-t_2)+\braket{\hat \hat \Psi(t_2)\Psi^{\dag}(t_1)\}}\Theta(t_2-t_1)$ and defining $\Theta(0)=1/2$ as it is done in ~\cite{stoof_field_1999}. However, what we do here is easier to generalize on averages of more operators., that follow next.

For the second order coherence function:
\begin{equation}
    g_2(0)=\frac{\braket{\hat \Psi^{\dag}\hat \Psi^{\dag}\hat \Psi\hat \Psi}}{\braket{\hat{N}}^2}
\end{equation}
we express the following average (here $\phi_i=\phi(t_i)$)
\begin{align}
    \braket{\ovl\psi_1\psi_2\ovl\psi_3\psi_4}{=}\braket{T_{C}\{\hat \Psi^{\dag}(t_1)\hat \Psi(t_2)\hat \Psi^{\dag}(t_3)\hat \Psi(t_4)\}}.
\end{align}
After inverting the Keldysh rotation:
\begin{align}
&\braket{\phi(t_{1})\overline{\phi(t_{2})}\phi(t_{3})\overline{\phi(t_{4})}}=\frac1{16}\langle\left[\psi(t^+_{1})+\psi(t^-_{1})\right]\times\nonumber\\
\times&\left[\overline{\psi}(t^+_{2})+\overline{\psi}(t^-_{2})\right]\left[\psi(t^+_{3})+\psi(t^-_{3})\right]\left[\overline{\psi}(t^+_{4})+\overline{\psi}(t^-_{4})\right]\rangle.
\end{align}
and making all the times to be equal, we obtain:
\begin{align}
    16\braket{|\phi|^4}{=}\braket{\hat \Psi_0\hat \Psi_0^{\dag}\hat \Psi_0^{\dag}\hat \Psi_0}{+}3\braket{\hat \Psi_0^{\dag}\hat \Psi_0\hat \Psi_0\hat \Psi_0^{\dag}}{+}4\braket{\hat \Psi_0\hat \Psi_0^{\dag}\hat \Psi_0\hat \Psi_0^{\dag}}{+}\nonumber\\
   {+}4\braket{\hat \Psi_0^{\dag}\hat \Psi_0\hat \Psi_0^{\dag}\hat \Psi_0}{+}2\braket{\hat \Psi_0^{\dag}\hat \Psi_0^{\dag}\hat \Psi_0\hat \Psi_0}{+}2\braket{\hat \Psi_0\hat \Psi_0\hat \Psi_0^{\dag}\hat \Psi_0^{\dag}}.
\end{align}

The use of equal-time commutation relations leads to
\begin{align}
    \braket{|\phi|^4}{=}\braket{\hat \Psi_0^{\dag}\hat \Psi_0^{\dag}\hat \Psi_0\hat \Psi_0}+\frac1{16}\left[32\braket{\hat{N_0}}+8\right].
\end{align}

That's all we need to construct $g_2(0)$:
\begin{equation}\label{eq:g2_general_calc}
    g_2(0)=\frac{{\braket{|\phi|^4}}- 2\braket{|\phi|^2}+\frac12}{\left(\braket{|\psi|^2}-\frac12\right)^2}=2+\frac{\braket{|\phi|^4}-2\braket{|\phi|^2}^2}{\left(\braket{|\phi|^2}-\frac12\right)^2}.
\end{equation}

Remembering that in the path-integral formalism that we use, $\phi$ denotes the corresponding coherent state, we get the well known limiting cases for Gaussian distribution (the nominator of the second term vanishes and we get $g_2(0)=2$) and a coherent state with $\braket{|\psi|^2}=|\braket{\phi}|^2$ and $\braket{|\phi_0|^4}=\braket{|\phi_0|^2}^2$ which leads to $g_2(0)=1+1/N^2$.  The latter converges to unity for high enough occupation.

\end{document}